**Current Switching of Topological Spin Chirality in the van der Waals Antiferromagnet $Co_{1/3}TaS_2$**

*Kai-Xuan Zhang,* Seungbok Lee, Woonghee Cho, and Je-Geun Park**

Dr. Kai-Xuan Zhang, Seungbok Lee, Woonghee Cho, Prof. Je-Geun Park
Department of Physics and Astronomy, Seoul National University, Seoul 08826, South Korea
E-mail: kxzhang.research@gmail.com; jgpark10@snu.ac.kr

Dr. Kai-Xuan Zhang, Seungbok Lee, Woonghee Cho, Prof. Je-Geun Park
Center for Quantum Materials, Department of Physics and Astronomy, Seoul National University, Seoul 08826, South Korea

Dr. Kai-Xuan Zhang, Seungbok Lee, Woonghee Cho, Prof. Je-Geun Park
Institute of Applied Physics, Seoul National University, Seoul 08826, South Korea

**Abstract:** Magnetic topology is central to modern quantum magnet, where spin chirality governs exotic spin winding, real-space Berry phase, and topological Hall effect. A key unresolved challenge is how to electrically switch topological spin chirality and its associated gauge flux, an essential requirement for manipulating its topological quantum properties. In this work, we propose and experimentally demonstrate the concept of current-switching spin chirality. We identify the new vdW antiferromagnet $Co_{1/3}TaS_2$ as an ideal platform, hosting a topological 3Q state with a minimum chirality cell, an ultrahigh skyrmion density, a non-centrosymmetric geometry, and a strong Berry curvature. We discover intrinsic self-torque-induced chirality switching within $Co_{1/3}TaS_2$, driven purely by current, without the need of heavy metals or a magnetic field, and with high energy efficiency. Our results establish a promising framework for electrically generating and controlling topological spin chirality, and demonstrate a practical route toward chiral spintronics. They can be naturally generalised to other skyrmion systems, offering new opportunities in symmetry control, topological manipulation, and spin-chirality–based quantum functionalities.

**Main Text:** Spin chirality naturally emerges in noncoplanar spin texture and represents one of the most intriguing quantum characteristics of magnetic topology[1,2]. In the canonical three-spin model, the scalar spin chirality is defined as $\chi = \langle \boldsymbol{S_1} \cdot (\boldsymbol{S_2} \times \boldsymbol{S_3}) \rangle$, where $\boldsymbol{S_1}$, $\boldsymbol{S_2}$, and $\boldsymbol{S_3}$ denote spins located on a triangular plaquette[1,2]. When an electron moves through such a topological spin configuration, it acquires a real-space Berry phase, giving rise to a variety of topological quantum phenomena, most notably the topological Hall effect[1,2]. In this sense, spin chirality acts as an emergent magnetic field, an effective gauge flux, whose sign determines the handedness of the spin texture and its underlying magnetic topology. Because this gauge flux directly couples to charge transport, the ability to create, manipulate, and ultimately switch spin chirality has become a central goal in condensed-matter physics and chiral spintronics. Despite its fundamental importance, however, a major open challenge remains: how to electrically reverse topological spin chirality between its two time-reversed states in a controlled and energy-efficient manner. This unresolved question motivates the present work.

Two-dimensional (2D) magnetic van der Waals (vdW) materials[3-9] have rapidly emerged as a fertile platform for exploring unconventional magnetism and topology. Among them, $Co_{1/3}TaS_2$ stands out as a new antiferromagnetic metal with unique magnetic topology[10,11]. It hosts a topological 3Q noncoplanar state stabilised by 3/4-filled Fermi surface nesting[10,11], forming a tetrahedral four-spin lattice that yields the highest skyrmion density known in vdW magnets[10,12]. Within this lattice, every three-spin plaquette carries a scalar spin chirality of the same sign. As a result, the effective chirality does not cancel. Instead, it remains a robust, uniform gauge flux throughout the long-period crystal, generating a fictitious magnetic field and a pronounced topological Hall effect[10,11,13]. Such a topological 3Q state[10,11] refers to the spin structure with uniform spin scalar chirality for triangular lattice planes. This state arises from the superposition of three distinct modulation vectors, Q1 = (1/2, 0, 0), Q2 = (0, 1/2, 0) and Q3 = (1/2, -1/2, 0) within a single domain. Co intercalation thus generates an atomic-scale spin-chirality platform ideally suited for the study of magnetic topology. Our recent work[13] demonstrated that ionic gating can control this chirality by tuning the Fermi level and nesting conditions, successfully reducing or weakening the topological Hall response[13]. However, such gating cannot reverse the chirality or toggle between its two time-reversed states. This limitation poses a critical open question: can the topological spin chirality in $Co_{1/3}TaS_2$ be electrically switched back and forth in a controlled manner? Addressing this question is essential for any practical exploitation of spin chirality in chiral spintronics.

We can outline our conceptual target into two parts: how to control and switch the complex spin arrangements of an exotic antiferromagnet. Antiferromagnetic spintronics offers several advantages: zero stray field, faster dynamics, and stability against magnetic perturbations, which enable higher density, scalability, and speed for next-generation memory devices and in-memory computing[14,15]. However, due to the strong internal field and negligible coupling to external perturbations, controlling an antiferromagnetic state remains an experimental challenge despite significant effort in the field.

Spin-orbit torque (SOT) provides powerful routes for manipulating spin structures using electrical current[16]. In the conventional approach, SOT arises in magnet/heavy-metal heterostructures, where the spin Hall effect in the heavy metal injects a spin current and exerts a torque on the adjacent magnet[16]. Such interfacial SOT can switch magnetic states either with the assistance of an in-plane magnetic field[17,18] or, more recently, without an external magnetic field through engineered symmetry breaking[19]. Beyond this conventional architecture, a distinct and rapidly developing mechanism, intrinsic self-SOT, has been identified in vdW magnets. Here, the torque originates within the magnet itself, without requiring a separate spin-source layer of heavy metal[20]. A prominent example is $Fe_3GeTe_2$, where the combination of nontrivial geometry, strong Berry curvature, and topological electronic bands produces a giant intrinsic SOT[20,21]. Additional inversion-symmetry breaking[22] further enhances this effect. Such an SOT effect enables efficient multi-level current switching[23] and even practical 3-terminal SOT-MRAM[24]. Subsequent studies have confirmed and expanded self-SOT behavior in $Fe_3GeTe_2$[25,26], and in related vdW magnets such as $Fe_3GaTe_2$[27-29] and $Fe_{0.5}Co_{0.5}GeTe_2$[30].

Importantly, $Co_{1/3}TaS_2$ also possesses a non-centrosymmetric, mirror-symmetry-broken space group and hosts possible topological bands[31]. These characteristics suggest the possibility of intrinsic self-SOT, making $Co_{1/3}TaS_2$ a compelling candidate for possible current-driven self-switching of its topological spin chirality without the need for external fields or heavy-metal layers. We emphasise that the above SOT and intrinsic self-SOT are frequently used in uniform ferromagnets to switch magnetic moments coherently. But this approach cannot be easily extended to an antiferromagnet of a complicated multi-spin structure. All the SOT arguments are based on extensive experience with SOT in uniform ferromagnets (e.g., the $Fe_3GeTe_2$ family). Thus, it is not *a priori* obvious before our real experimental demonstration that current-driven SOT can switch the magnetic moments of a complex antiferromagnet like $Co_{1/3}TaS_2$. We note that to our best knowledge, the combination of self-SOT and field-free switching has not been demonstrated for a vdW ferromagnet, let alone a complicated antiferromagnet.

In this work, we demonstrate electrical switching of topological spin chirality using the layered antiferromagnet $Co_{1/3}TaS_2$ as a model system. We first establish the general concept of switching between time-reversal pairs of spin chirality by electrical current and clarify why the topological 3Q state of $Co_{1/3}TaS_2$ provides the essential ingredients for this functionality. Interestingly, the topological spin chirality can be electrically switched by a distinct intrinsic self-SOT within $Co_{1/3}TaS_2$ itself: it can be switched purely by current, without an external magnetic field and with high energy efficiency. This discovery of intrinsic self-SOT with magnetic-field-free switching in an antiferromagnet establishes a comprehensive framework for current-driven manipulation of topological spin chirality, validating our proposed chirality-switching concept. As such, we demonstrate that the quantum antiferromagnet $Co_{1/3}TaS_2$ can be fully controlled by electrical means, opening up new opportunities in chiral spintronics and magnetic topology.

***Concept of switching spin chirality*:** We begin by elaborating on the fundamental idea of electrically switching spin chirality. As illustrated in **Figure 1**, a noncoplanar spin texture, such as the canonical three-spin configuration on a triangular plaquette, carries a finite topological spin chirality, defined as $\chi = \langle \boldsymbol{S_1} \cdot (\boldsymbol{S_2} \times \boldsymbol{S_3}) \rangle$, where $\boldsymbol{S_1}$, $\boldsymbol{S_2}$, and $\boldsymbol{S_3}$ denote the spins at the vertices of the triangular plaquette[1,2]. This scalar chirality acts as a measure of the solid angle subtended by the three spins.

When an electron moves in the vicinity of such a noncoplanar texture, it acquires an additional real-space Berry phase that effectively behaves like a fictitious magnetic field or quantum gauge flux. This quantum gauge flux deflects the electron trajectory and produces an additional Hall contribution, known as the topological Hall effect[1,2]. In Fig. 1, the handedness of the spin configuration (chirality) is represented by the left-circular arrow, while the up-straight arrow illustrates the associated gauge flux. This intrinsic correspondence between spin chirality and emergent gauge flux underlies our central idea: current-driven control of the chirality state, i.e., switching between its time-reversal counterparts by electrically manipulating the underlying spin texture.

Under a time-reversal symmetry operation, all three spins invert their orientations and the scalar spin chirality $\chi = \langle \boldsymbol{S_1} \cdot (\boldsymbol{S_2} \times \boldsymbol{S_3}) \rangle$ changes its sign accordingly. This corresponds to switching from one handedness of the noncoplanar texture to its time-reversed counterpart, depicted by the right-circular chirality and the downward emergent gauge flux. Because the fictitious gauge field directly reflects the sign of the real-space Berry phase, it also reverses under time-reversal operation, consistent with how an ordinary magnetic field transforms.

Our focus is on achieving this transformation electrically, i.e., switching between the two chirality states and their opposite gauge fluxes by applying a current. As schematically illustrated in the upper and lower panels of Fig. 1, the goal is to drive the system between the two time-reversal partners of spin chirality using current-induced torque. Achieving such current-driven chirality reversal provides a direct mechanism to control magnetic topology and the associated emergent electrodynamics.

***Spin chirality in topological 3Q states of vdW quantum antiferromagnet $Co_{1/3}TaS_2$:*** To pursue electrical switching of spin chirality, we employ the vdW antiferromagnet $Co_{1/3}TaS_2$ as our model system because its symmetry, topology, and electronic structure align exceptionally well with the requirement for current-driven control. As shown in **Figure 2a**, $Co_{1/3}TaS_2$ crystallises in a hexagonal, non-centrosymmetric structure with space group $P6_322$[10]. Intercalation of Co atoms into the vdW gap of $TaS_2$ lowers the original centrosymmetric space group $P6_3/mmc$ of pristine 2H-$TaS_2$ to this non-centrosymmetric one, simultaneously removing all mirror symmetries[10].

The combined breaking of inversion and mirror symmetries is crucial, as it enables the generation of current-driven SOT within the material. Such SOTs enable manipulation and switching of the underlying spin texture- either with an in-plane magnetic field[17,32] or, in suitably engineered systems, even without a magnetic field[33,34]. These symmetry conditions make $Co_{1/3}TaS_2$ an ideal candidate for investigating the possibility of electrically controlling topological spin chirality in a quantum antiferromagnet.

Moreover, the non-symmorphic symmetry of $6_3$ screw symmetry in $Co_{1/3}TaS_2$ enlarges the primitive unit cell and consequently induces Brillouin zone folding. This zone folding forces electronic bands to intersect or hybridise, particularly along high-symmetry directions, thereby increasing the likelihood of band crossings and symmetry-protected degeneracies. As suggested in previous work[31], these band crossings provide beneficial conditions for generating topological electronic states accompanied by substantial Berry curvature hotspots.

When combined with the material's non-centrosymmetric crystal geometry, this topological electronic structure creates a favorable environment for possibly bearing intrinsic self-SOT, analogous to that observed in vdW ferromagnet $Fe_3GeTe_2$ and its related compounds[20-30]. However, in sharp contrast, we would like to emphasise that intrinsic self-SOT switching has never been demonstrated in an antiferromagnet or in the absence of a magnetic field to date. In this picture, the Berry curvature acts as an internal source of spin current, enabling current-driven torque without relying on an external heavy-metal layer. Thus, the symmetry-enforced band topology of $Co_{1/3}TaS_2$ provides a promising pathway for pursuing a novel intrinsic self-SOT mechanism capable of manipulating its topological spin chirality.

Beyond its favorable symmetry and electronic characteristics for current-driven SOT, intercalation of Co atoms also gives rise to an exotic noncoplanar spin texture in the layered antiferromagnetic state of $Co_{1/3}TaS_2$. As shown in Figure 2b, the system hosts a topological tetrahedral 3Q state, stabilised by its 3/4-filled Fermi surface nesting[10,11,13]. The resulting four-spin lattice is composed of two triangular cells, each carrying a scalar spin chirality of the same sign and therefore generating an emergent gauge flux in the same direction, illustrated by the left-circular and up-straight arrows. Consequently, the minimum four-spin unit already exhibits a finite spin chirality, and this nonzero chirality persists throughout the long-period crystal. This microscopic mechanism underlies the pronounced topological Hall effect observed in

$Co_{1/3}TaS_2$[10,11,13].

Figure 2c presents the two possible 3Q spin configurations: one with positive spin chirality and another with its time-reversal partner bearing negative spin chirality. These two states can be interconverted by applying an out-of-plane magnetic field $H_z$, which reverses the emergent gauge flux and produces the characteristic Hall hysteresis loop shown in Figure 2d. This pair of chirality states provides the essential two-level system that we seek to switch electrically using current-driven torques.

Therefore, Co intercalation simultaneously engineers the crystal symmetry, electronic topology, and topological spin texture of $Co_{1/3}TaS_2$, producing a non-centrosymmetric structure, Berry curvature-rich bands, and a robust 3Q chirality lattice. These combined consequences uniquely position $Co_{1/3}TaS_2$ as an ideal platform for examining our central concept of "current-switching spin chirality".

***Spin chirality self-switching by pure current in $Co_{1/3}TaS_2$ itself:*** We now investigate whether the spin chirality of $Co_{1/3}TaS_2$ can be switched through a distinct intrinsic mechanism, a self-SOT generated within the material itself. As discussed above, Co intercalation gives rise to a unique combination of non-centrosymmetric crystal geometry, band topology, and strong Berry curvature, alongside its topological 3Q magnetism. These ingredients together favor the emergence of intrinsic self-SOT, analogous to the giant self-SOT discovered in vdW magnet $Fe_3GeTe_2$[20-22]. Furthermore, the reduced symmetry and the absence of mirror planes in $Co_{1/3}TaS_2$ create conditions that allow magnetic-field-free switching, similar to the symmetry-enabled SOT reversal demonstrated in $WTe_2$[33].

To test this possibility on such a complex antiferromagnet, we perform the standard chirality-switching experiment but with two critical advances: (1) no heavy-metal layers, ensuring that any torque originates solely from $Co_{1/3}TaS_2$ itself, and (2) no magnetic field, so that any observed switching reflects purely intrinsic symmetry-driven self-SOT. This new configuration enables us to directly investigate whether $Co_{1/3}TaS_2$ can spontaneously generate sufficient spin-orbit torque to self-switch its topological spin chirality.

**Figure 3a** illustrates the optical image of the $Co_{1/3}TaS_2$ device and the measurement scheme. A large writing current is first applied to induce SOT-mediated switching of the spin chirality. After the writing pulse is removed, a small reading current is used to measure the Hall resistance $R_{xy}$, which reflects the nonvolatile chiral state established by the preceding pulse. Figure 3b shows the temperature-dependent longitudinal resistance ($R_{xx}$-$T$), which exhibits metallic behavior over the measured range. The transverse resistance ($R_{xy}$-$H_z$) curves in Fig. 3c reveal the characteristic topological Hall effect of $Co_{1/3}TaS_2$, which appears below the Neel temperature $T_N$ ~25 K. These measurements confirm that the $Co_{1/3}TaS_2$ nanoflake preserves the intrinsic topological 3Q state and its finite spin chirality.

Joule heating is an unavoidable yet important issue in current-related experiments, including current-driven SOT switching[18,20]. We have taken extreme care to quantify the Joule heating effects of the device. As shown in Fig. S4, we have recorded the $R_{xx}$-$T$ curve (Fig. S4a) and $R_{xx}$-$I$ curve measured at 15 K (Fig. S4b), from which we notice that a current of 3.7 mA can increase the temperature maximum up to $T_N$ of 25 K (Fig. S4c). Therefore, applying a current smaller than 3.7 mA will guarantee the temperature remaining below the $T_N$ during the measurements (Fig. S4c). Based on this calibration, we performed the magnetic-field-free switching experiments with writing currents up to 3.3 mA.

Crucially, nonvolatile electrical switching of this chirality is demonstrated in Figure 3d. Without a magnetic field, the Hall resistance exhibits a clear current-induced hysteresis: applying writing current pulses toggles $R_{xy}$ between two stable values corresponding to opposite chirality states. This unambiguously shows that the current-driven self-SOT can reverse the topological spin chirality in the pristine $Co_{1/3}TaS_2$ device.

Figure 3e summarises magnetic-field-free switching at different temperatures. Hall resistance switching is reproduced and observed only below $T_N$, and vanishes near or above $T_N$, confirming that the effect originates from spin-texture modification and chirality reversal, rather than extrinsic artifacts.

We provide a more quantitative description of the switching using the $R_{xy}$ switching ratio: In Fig. 3d,e, at nominal temperature 15 K, the $R_{xy}$ switching ends with a current of 2.4-2.7 mA. Due to Joule heating, the estimated sample temperature at this current is approximately 20 K. The observed $R_{xy}$ switching magnitude is about 0.01 Ω. For comparison, the zero-field Hall resistance shown in Fig. 3c is about 0.04 Ω at 15 K and 0.012 Ω at 20 K, respectively. Therefore, the corresponding $R_{xy}$ switching ratio is between 0.01/0.04 = 25% and 0.01/0.012 = 83.3%.

Together, these results provide direct experimental validation of our central concept: current-controlled switching of topological spin chirality in the unique quantum magnet $Co_{1/3}TaS_2$.

We reproduce the current-driven self-switching of topological spin chirality in another $Co_{1/3}TaS_2$ device without a magnetic field. **Figure 4a** shows the optical image of another $Co_{1/3}TaS_2$ device fabricated without a heavy-metal layer, and Figure 4b displays its temperature-dependent longitudinal resistivity ($\rho_{xx}$-$T$), confirming metallic behaviour. The $R_{xy}$-$H_z$ curves in Figure 4c exhibit a clear topological Hall hysteresis below the Neel temperature $T_N$ ~25 K, demonstrating that the exfoliated flakes retain the intrinsic 3Q chiral antiferromagnetic state.

Similarly to the former example, when we apply a writing current pulse, the Hall resistance shows robust current-induced hysteresis (Fig. 4d), indicating that the spin chirality of $Co_{1/3}TaS_2$ is switched purely by intrinsic self-SOT, without the need for a heavy-metal layer or external magnetic field. Such magnetic-field-free switching is absent at temperatures above $T_N$,, e.g., 40 K (Fig. 4e), thereby excluding extrinsic artifacts and supporting the magnetic origin again.

A passing remark, the $R_{xy}$ dips and peaks near the switching current in Fig. 3d,e reflect the $R_{xy}$ fluctuations, which are frequently observed in current-switching experiments[32,35,36]. These fluctuations likely arise from the nucleation and annihilation of magnetic (or equivalent spin-chirality) domains, causing the Hall signal to transition between metastable states. Specifically, the application of a large pulse current can generate local variations in magnetic properties, including multiple-domain nucleation and domain-wall propagation. SOT is known to favor such complex domain nucleation and motion. Consequently, all these domain dynamics can lead to non-smooth transition and the observed $R_{xy}$ fluctuations. In addition, the step-like feature during current-driven switching in Figs. 3e and 4d also reflect the domain dynamics and pinning in $Co_{1/3}TaS_2$'s topological 3Q state. During current-driven switching, if discrete domain rearrangements or multi-stage transitions occur in the Hall signal path, the signal exhibits a step-like feature; conversely, more collective domain dynamics yields a smoother switching curve. These results suggest that studying these domain dynamics would be an interesting direction. However, this promising direction is very challenging at the moment, especially for small antiferromagnetic domains, which are beyond the scope of the present work and warrant future studies.

The critical switching current density is approximately $1.8\times10^6$ A/cm², comparable to or even lower than values reported in recent highly efficient SOT works[33,37]. These results therefore establish that $Co_{1/3}TaS_2$ can autonomously generate sufficient SOT to switch its topological spin chirality, achieving field-free and energy-efficient operation. This self-SOT pathway offers a fundamentally new mechanism for controlling magnetic topology in vdW antiferromagnets. We have summarised the novelty, challenges, unique features, and significance of the present work exclusively in the Supporting Information (see Supporting Note 1 for more details).

***Conclusion:*** In summary, we have introduced and experimentally demonstrated the concept of current-switching topological spin chirality using the vdW quantum antiferromagnet $Co_{1/3}TaS_2$. Taking full advantage of the material's topological 3Q states and its atomic-scale chirality lattice, we propose and exploit a distinct intrinsic self-SOT mechanism, enabling field-free and heavy-metal-free electrical switching of spin chirality within $Co_{1/3}TaS_2$ itself. Eventually, we realise the clear, nonvolatile, reversible, and highly energy-efficient switching of the topological spin chirality via pure current, by combining intrinsic self-SOT and magnetic-field-free switching in an exotic vdW antiferromagnet. These results collectively establish a comprehensive framework for electrically creating, controlling, and manipulating topological spin chirality. Our findings reveal a new dimension of chiral spintronics and provide a versatile strategy applicable to other skyrmion systems, noncoplanar magnets, and chirality-related quantum effects.

**Experimental Section**

*Single crystal synthesis:* $Co_{1/3}TaS_2$ single crystals were grown by a two-step growth method. To ensure compositional homogeneity, a precursor was prepared via a solid-state reaction. A well-ground mixture of Co (Alfa Aesar, >99.99%), Ta (Sigma Aldrich, >99.99%), and S (Sigma Aldrich, >99.99%) was sealed in an evacuated quartz ampoule and sintered at 900 °C for 10 days. The resulting polycrystalline precursor and $I_2$ transport agent (4.5 mg $I_2/cm^3$) were then placed in an evacuated quartz tube, and single crystals were grown by the chemical vapor transport (CVT) method. The quartz tube was heated in a two-zone furnace with a temperature gradient from 960°C to 840 °C for 2 weeks. The obtained single crystals were characterised by X-ray diffraction (XRD) and Raman spectroscopy, as shown in Figures S1 and S2.

*Device fabrication:* $Co_{1/3}TaS_2$ nanoflakes were mechanically exfoliated onto a $SiO_2$/Si wafer, and a suitable sample was chosen for transport investigations. The polymethyl methacrylate (PMMA) A7 was spin-coated onto the nanoflake at a rate of 4000 rpm, and then post-baked at a mild temperature of 130 ºC for 1.5 mins. Afterwards, the electrodes were designed using the electron beam lithography, and then 80/10 nm Au/Ti metals were deposited in order by the electron beam evaporator.

*Electrical Transport measurement:* Electrical transport measurements were performed using a home-built closed-circuit resistance setup and a commercial cryogenic system. Those measurements were carried out using a Keithley 6220, a Keithley 2182, and a lock-in amplifier.

The $R_{xy}$-$H_z$ curves were measured with an applied out-of-plane magnetic field. The writing current pulse was applied for 1 s, and the Hall signal was measured after a delay of 10 s.

**Supporting Information**

Supporting Information is available from the Wiley Online Library or the author.

**Acknowledgments**

K.-X.Z. and S.L. contributed equally to this work. We acknowledge Jihoon Keum for the helpful discussions and experimental support. The work at CQM and SNU was supported by the Samsung Science & Technology Foundation (Grant No. SSTF-BA2101-05) and the Leading Researcher Program of the National Research Foundation of Korea (Grant No. RS-2020-NR049405).

Received: ((will be filled in by the editorial staff))
Revised: ((will be filled in by the editorial staff))
Published online: ((will be filled in by the editorial staff))

**Figures**

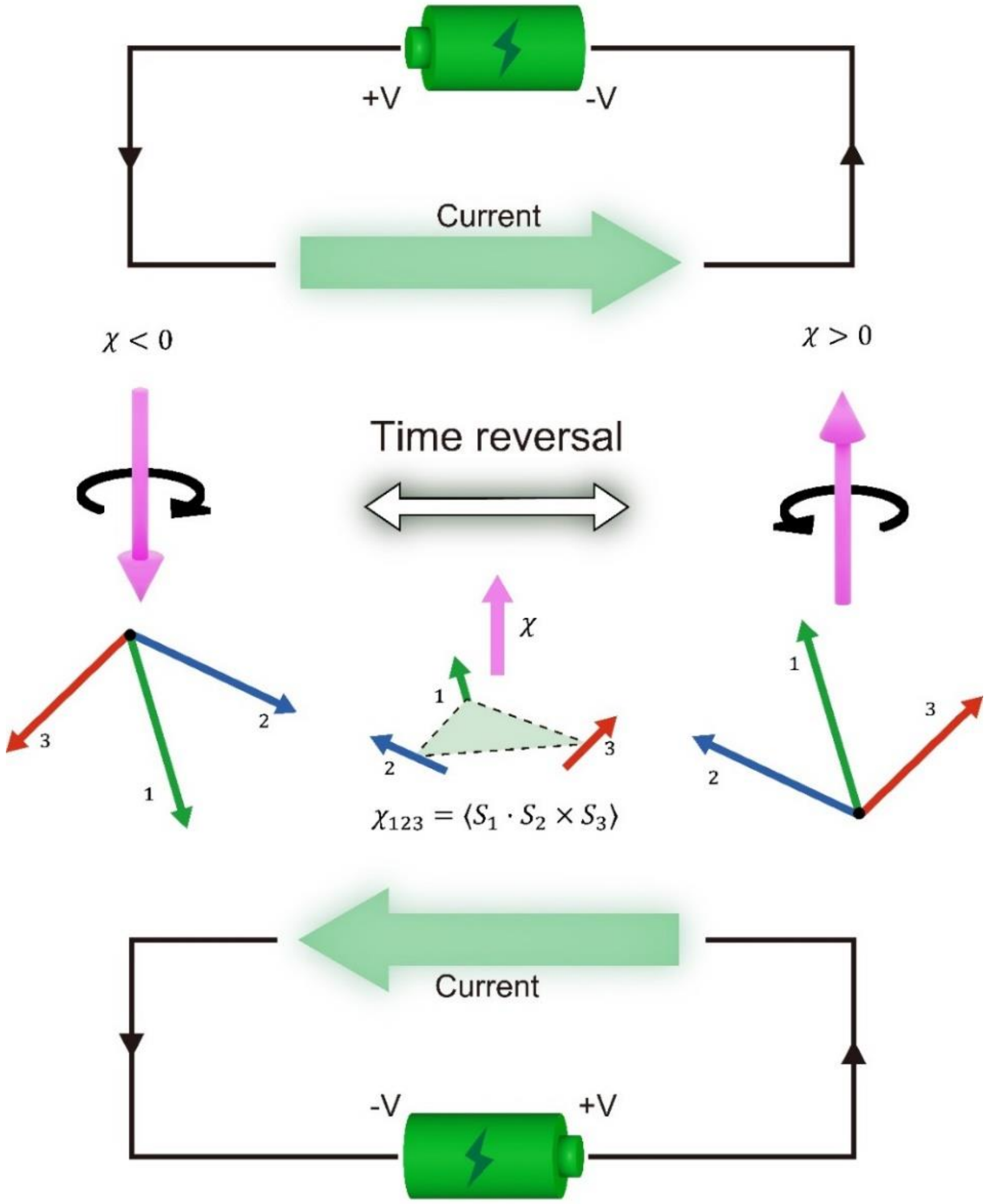

**Figure 1.** Concept of current-switching spin chirality and its associated gauge flux. In a noncoplanar three-spin configuration, spin chirality can arise as $\chi = \langle \boldsymbol{S_1} \cdot (\boldsymbol{S_2} \times \boldsymbol{S_3}) \rangle$, where $\boldsymbol{S_1}$, $\boldsymbol{S_2}$, and $\boldsymbol{S_3}$ represent spins at the vertices of a triangular plaquette. It can give rise to a real-space Berry phase to the electron's wave function and behaves like a fictitious magnetic field or gauge flux, generating the topological Hall effect. As illustrated, the spin texture determines both the chirality of spin and the direction of the gauge flux. Under a time-reversal operation, spin, chirality, and gauge flux all flip their directions simultaneously. For fundamental control and practical

application, we propose switching spin chirality via current, as highlighted by the thick green arrows.

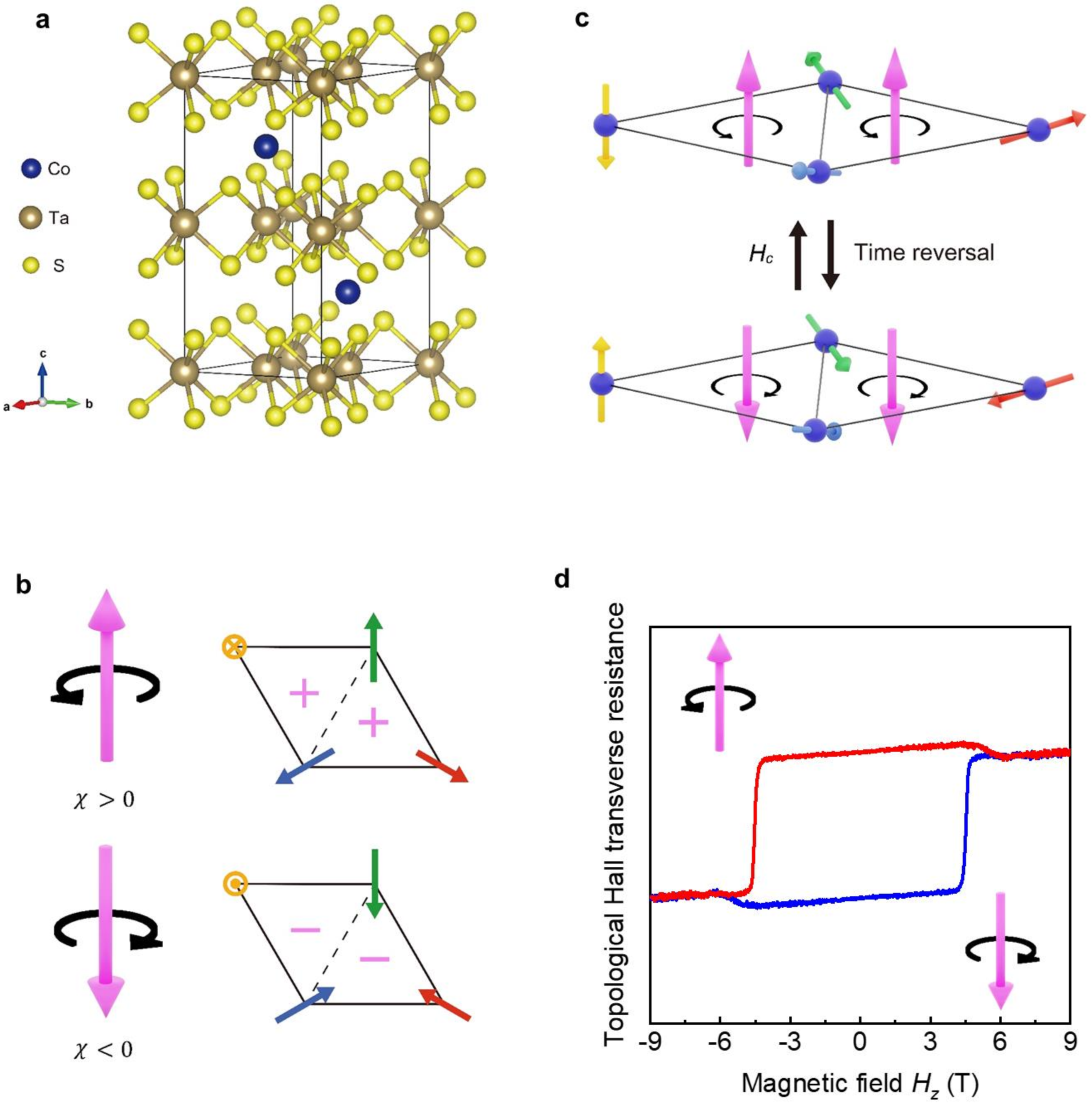

**Figure 2**. Unique spin chiral properties of $Co_{1/3}TaS_2$. a) Crystal structure of $Co_{1/3}TaS_2$. The black line indicates a unit cell, and the blue, brown, and yellow balls represent Co, Ta, and S atoms, respectively. Co atoms are intercalated into the vdW gap of the $TaS_2$ layer, making several important quantum consequences as numerated in the main text. b) Spin chirality of the topological tetrahedral 3Q states in $Co_{1/3}TaS_2$. For the four-spin primitive lattice, each three-spin cell hosts a spin chirality of the same sign, rendering a robust, nonzero spin chirality throughout the whole crystal structure. c) Under time-reversal symmetry operation, e.g., by out-of-plane magnetic field

$H_z$, each spin flips in the four-spin lattice, so do the spin chirality and gauge flux. d) Experimental outcome of (c): hysteresis loop of the topological Hall effect by sweeping $H_z$.

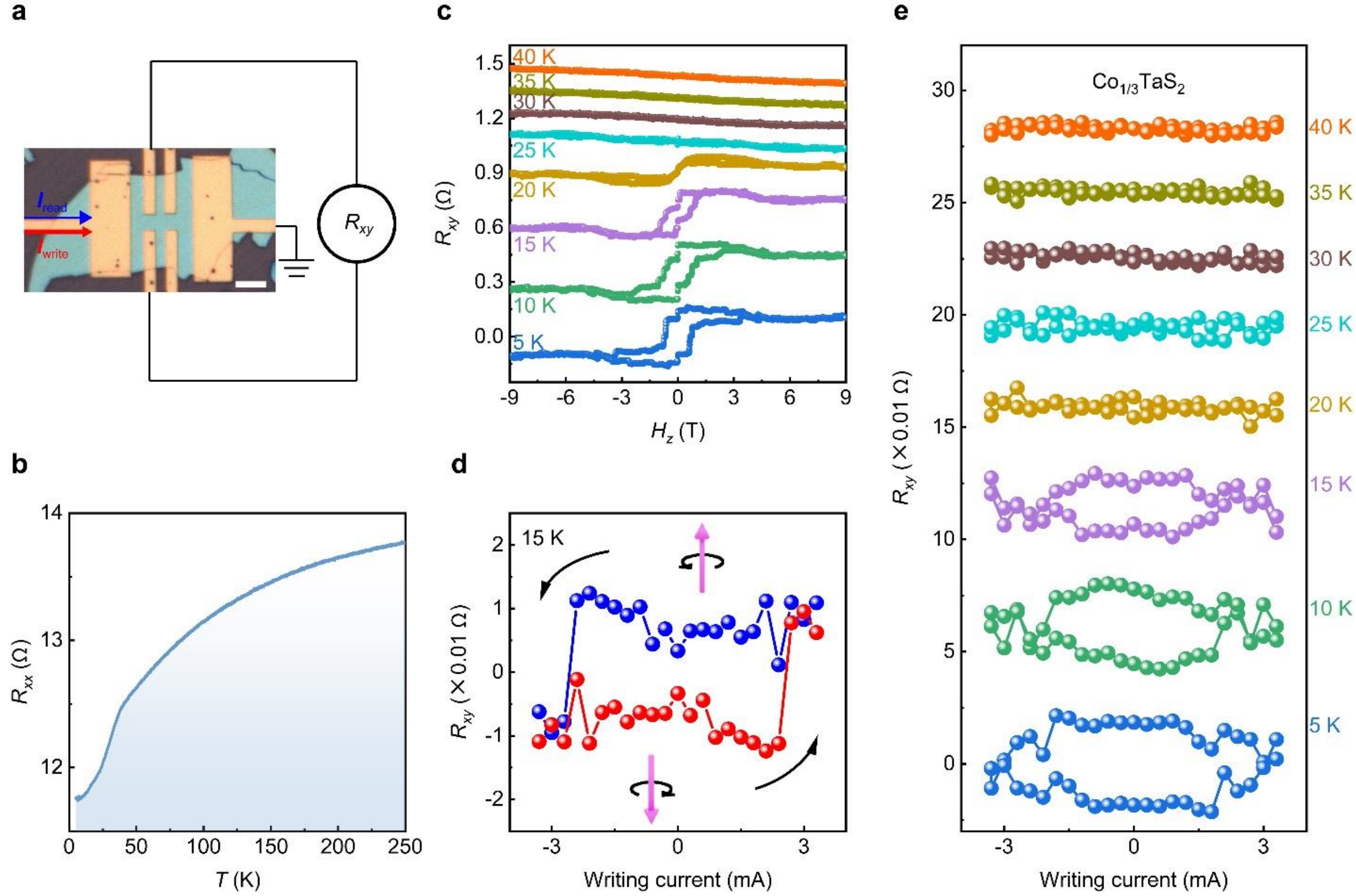

**Figure 3.** Intrinsic current switching of spin chirality in $Co_{1/3}TaS_2$. a) Optical image of the pristine $Co_{1/3}TaS_2$ device and the schematic of the switching measurement. The white scale bar represents 5 μm. b) $R_{xx}$-$T$ curve of the device, showing metallic behavior. c) $R_{xy}$-$H_z$ curves at various temperatures, where clear hysteresis loops emerge below the Neel temperature of $T_N$~25 K. d) Demonstration of spin chirality switching purely by current in the pristine $Co_{1/3}TaS_2$ device. e) Temperature dependence of the switching behavior. The current-driven spin chirality switching emerges below the Neel temperature of $T_N$~25 K, and disappears near or above $T_N$, confirming its magnetic origin.

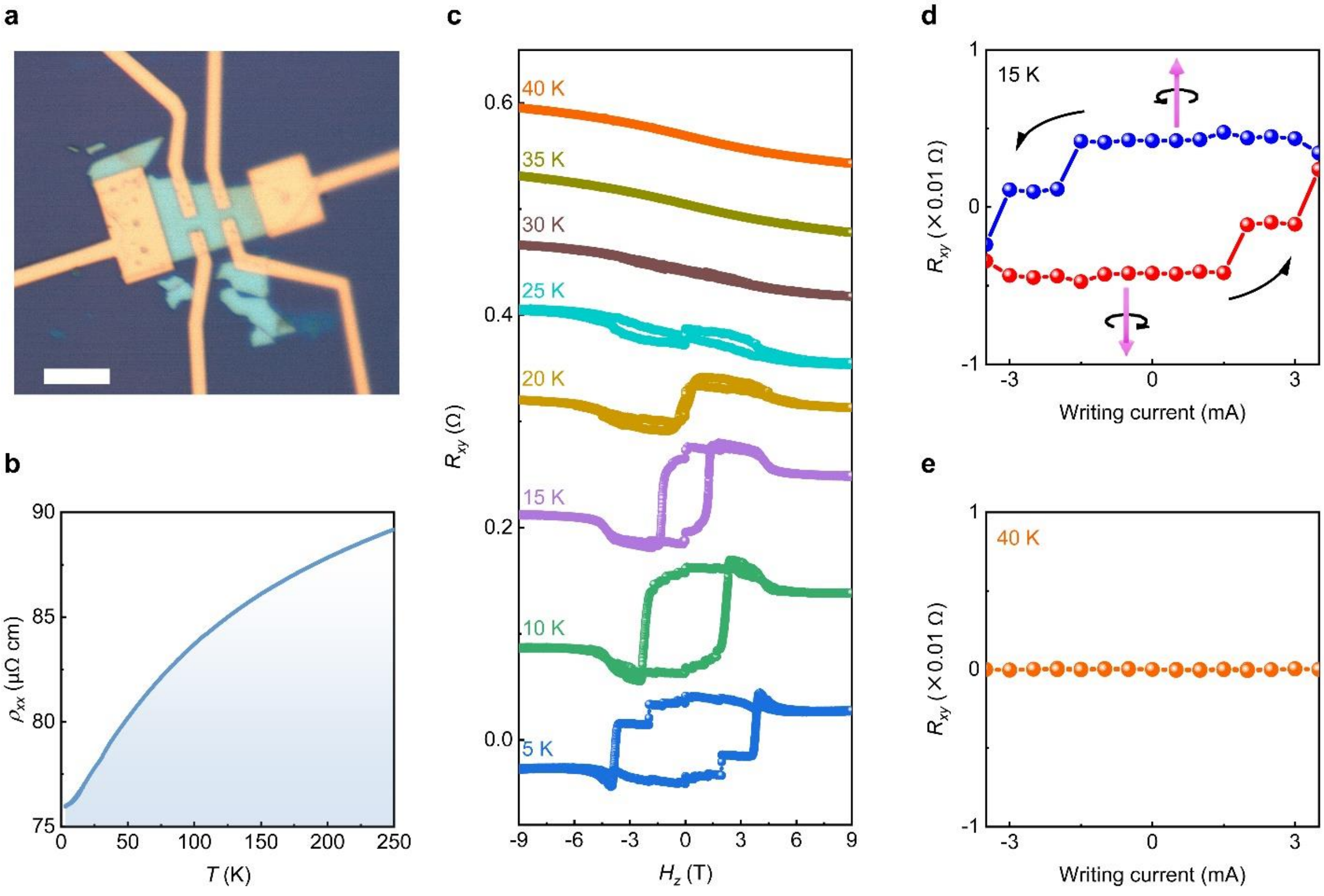

**Figure 4.** Current of switching spin chirality in another pristine $Co_{1/3}TaS_2$ device. a) Optical image of the pristine $Co_{1/3}TaS_2$ device. The white scale bar represents 10 μm. b) $\rho_{xx}$-$T$ curve of the device, showing metallic behavior. c) $R_{xy}$-$H_z$ curves at various temperatures, where clear hysteresis loops emerge below the Neel temperature of $T_N$~25 K. d) Demonstration of spin chirality switching purely by current in the pristine $Co_{1/3}TaS_2$ device. e) $R_{xy}$ versus writing current at 40 K. No current-driven spin chirality switching is observed at a temperature above the Neel temperature of $T_N$~25 K, supporting its magnetic origin.

## The table of contents entry

Spin chirality plays a central role in quantum magnetism, governing spin winding and generating real-space Berry phases. We propose the intriguing concept of current-switching spin chirality and demonstrate it using the van der Waals antiferromagnet $Co_{1/3}TaS_2$ as an ideal testbed. We demonstrate chirality reversal via an unconventional current-driven intrinsic spin-orbit torque in pristine $Co_{1/3}TaS_2$, purely by electrical current and with high energy efficiency. This route establishes a new framework for creating, controlling, and exploiting topological spin chirality. They can be readily extended to other skyrmion-hosting or chirality-driven quantum materials. Our results exemplify electrical control of symmetry, topology, and chiral spintronics in a van der Waals quantum antiferromagnet.

Kai-Xuan Zhang,* Seungbok Lee, Woonghee Cho, and Je-Geun Park*

Title: Current Switching of Topological Spin Chirality in the van der Waals Antiferromagnet $Co_{1/3}TaS_2$

ToC figure

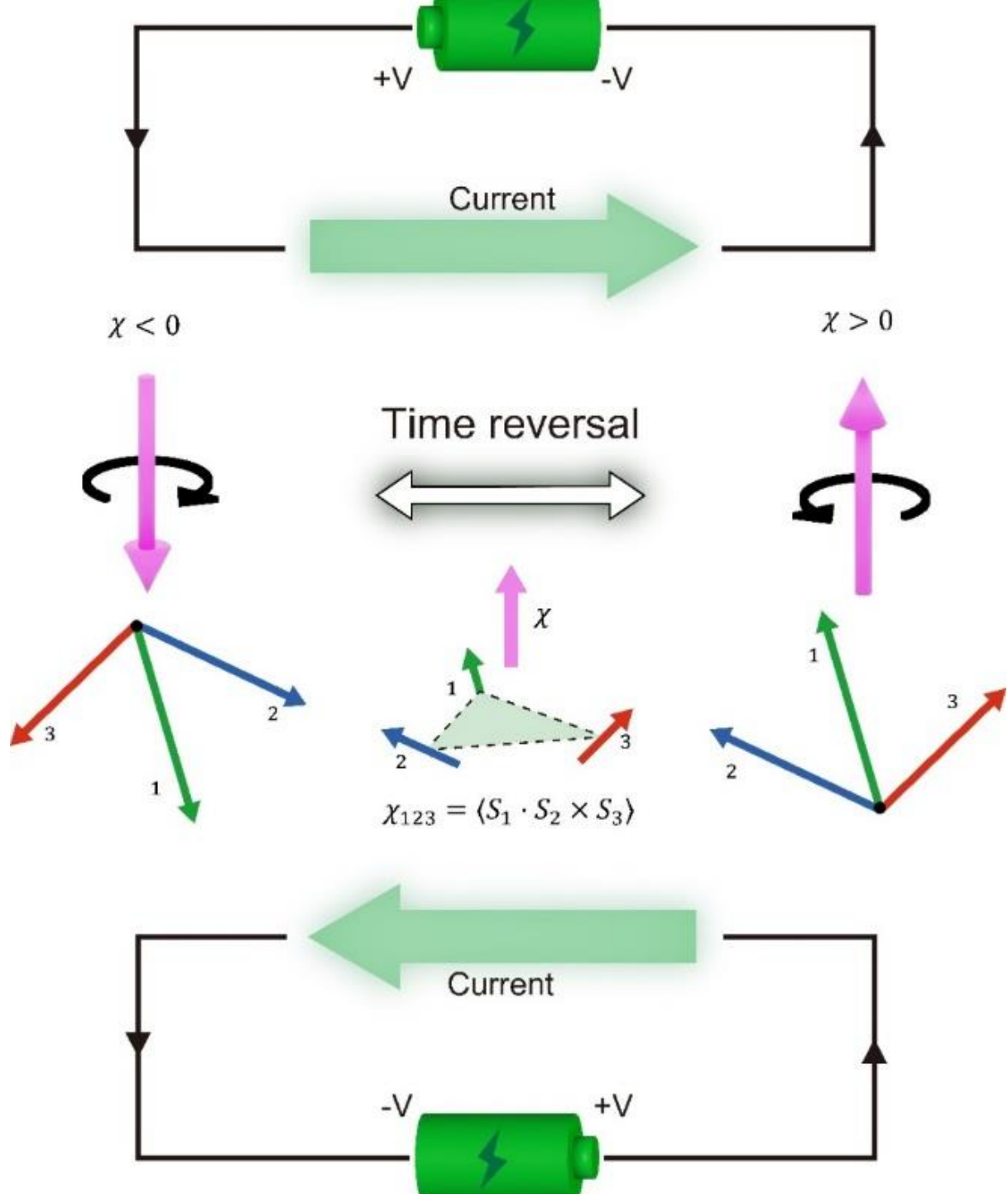

# Supporting Information

## Current Switching of Topological Spin Chirality in the van der Waals Antiferromagnet $Co_{1/3}TaS_2$

*Kai-Xuan Zhang,* Seungbok Lee, Woonghee Cho, and Je-Geun Park**

K.-X.Z. and S.L. contributed equally to this work.

**Supporting Note 1. The novelty, challenges, distinctions, and significance of the present work.**

For clarity, we discuss the novelty, challenges, distinctions, and significance of our work logically from our starting premise to the final discoveries:

1. **Novel focus on "switching spin chirality in exotic topological 3Q antiferromagnet":**

   Our primary object is to control and switch a *topological order parameter*—the scalar spin chirality (or equivalently, the emergent real-space gauge flux)—in an *exotic topological 3Q antiferromagnet*. This focus is novel and distinct from most SOT studies in the literature.

2. **The challenge of electrically switching an antiferromagnet:**

   The van der Waals $Co_{1/3}TaS_2$ antiferromagnet, which has attracted significant interest, hosts a topological spin chirality with unique magnetic topology. This multi-spin quantity can be measured via the topological Hall effect. Therefore, we propose the concept of current switching topological spin chirality based on $Co_{1/3}TaS_2$.

   Our experiment requires switching the spins, since the spin texture governs the spin chirality. Therefore, the concept target can be broken down into two parts: how to control and switch the complex spin arrangements of an exotic antiferromagnet. Antiferromagnetic spintronics offers several advantages: zero stray field, faster dynamics, and stability against magnetic perturbations, which enable higher density, scalability, speed, and stability for next-generation memory devices and in-memory computing. However, due to the strong internal field and negligible coupling to external perturbations, switching an antiferromagnet remains a significant challenge despite considerable efforts in the field.

3. **The challenge of predicting spin chirality switching in complex antiferromagnet $Co_{1/3}TaS_2$:**

To address the above challenge, we proposed current-driven spin-orbit torque (SOT) mechanism to switch the complex antiferromagnet $Co_{1/3}TaS_2$. It is worth noting that while SOT has often been used to switch magnetic moments coherently in uniform ferromagnets, extending this to an antiferromagnet with a complex multi-spin structure is not straightforward. For example, in the topological 3Q state of the exotic antiferromagnet $Co_{1/3}TaS_2$, local moments point along multiple noncoplanar directions; therefore, a current-induced spin accumulation or spin polarisation cannot trivially generate a uniform effective torque. Since existing SOT models are largely based on uniform ferromagnets (e.g., the $Fe_3GeTe_2$ family), it was not *a priori* obvious that current-driven SOT would successfully switch the magnetic moments of a complicated antiferromagnet. This open question motivated our experiment: Can an applied current achieve reproducible, non-volatile, and efficient magnetic moment switching, thereby producing equivalent spin-chirality switching? Before our real experimental demonstration, such a possibility had not been established for $Co_{1/3}TaS_2$.

4. **Novel and efficient "intrinsic-self-SOT + field-free switching" for antiferromagnet:**

In sharp contrast to conventional SOT induced by heavy metals such as Pt, our group previously demonstrated the intrinsic self-SOT in the vdW uniform ferromagnet $Fe_3GeTe_2$. However, that work and subsequent studies by others reported only the intrinsic self-SOT in ferromagnets, and the switching of magnetic moments by self-SOT typically requires an assisting in-plane or out-of-plane magnetic field.

Our present work on $Co_{1/3}TaS_2$ demonstrates, for the first time, the self-SOT in an exotic complex antiferromagnet. More importantly, the self-SOT switching (together with topological spin chirality switching) occurs without an external magnetic field in a single antiferromagnet. The self-SOT switching is non-volatile, reproducible, reversible, and highly energy-efficient: the switching current density ($1.8\times10^6$ A/cm²) is competitive with, or even lower than, the most efficient SOT systems reported to date. This highlights our contributions to the fields of SOT and antiferromagnetic spintronics, beyond the specific realisation of chirality switching.

## 5. Realisation of "current switching topological spin chirality":

Spin chirality and magnetic topology are fundamental concepts connecting these highly interesting topics: noncoplanar spin textures, real-space Berry phases, and the topological Hall effect. A longstanding challenge for both fundamental understanding and device application has been the electrical switching of topological spin chirality and its emergent gauge flux.

In this work, we introduce the concept of current-switching spin chirality and experimentally realise it using the vdW antiferromagnet $Co_{1/3}TaS_2$. We demonstrate that intrinsic SOT can electrically switch the topological spin chirality purely by current, without the need for heavy metals or a magnetic field. Our results exhibit clear, nonvolatile, reversible, and highly energy-efficient current-induced chirality switching, facilitated by its non-centrosymmetric lattice, Berry-curvature-rich electronic structure, and topological magnetism. This work establishes a novel framework for electrically generating, controlling, and exploiting topological spin chirality, which may apply to other skyrmion-hosting and chirality-driven systems.

In summary, our work provides a novel demonstration of the electrical writing of the chiral 3Q antiferromagnetic state in $Co_{1/3}TaS_2$. We believe it lays an important new foundation for future work combining vdW magnets, antiferromagnetic spintronics, SOT, and topological spin chirality.

**Supporting Figures**

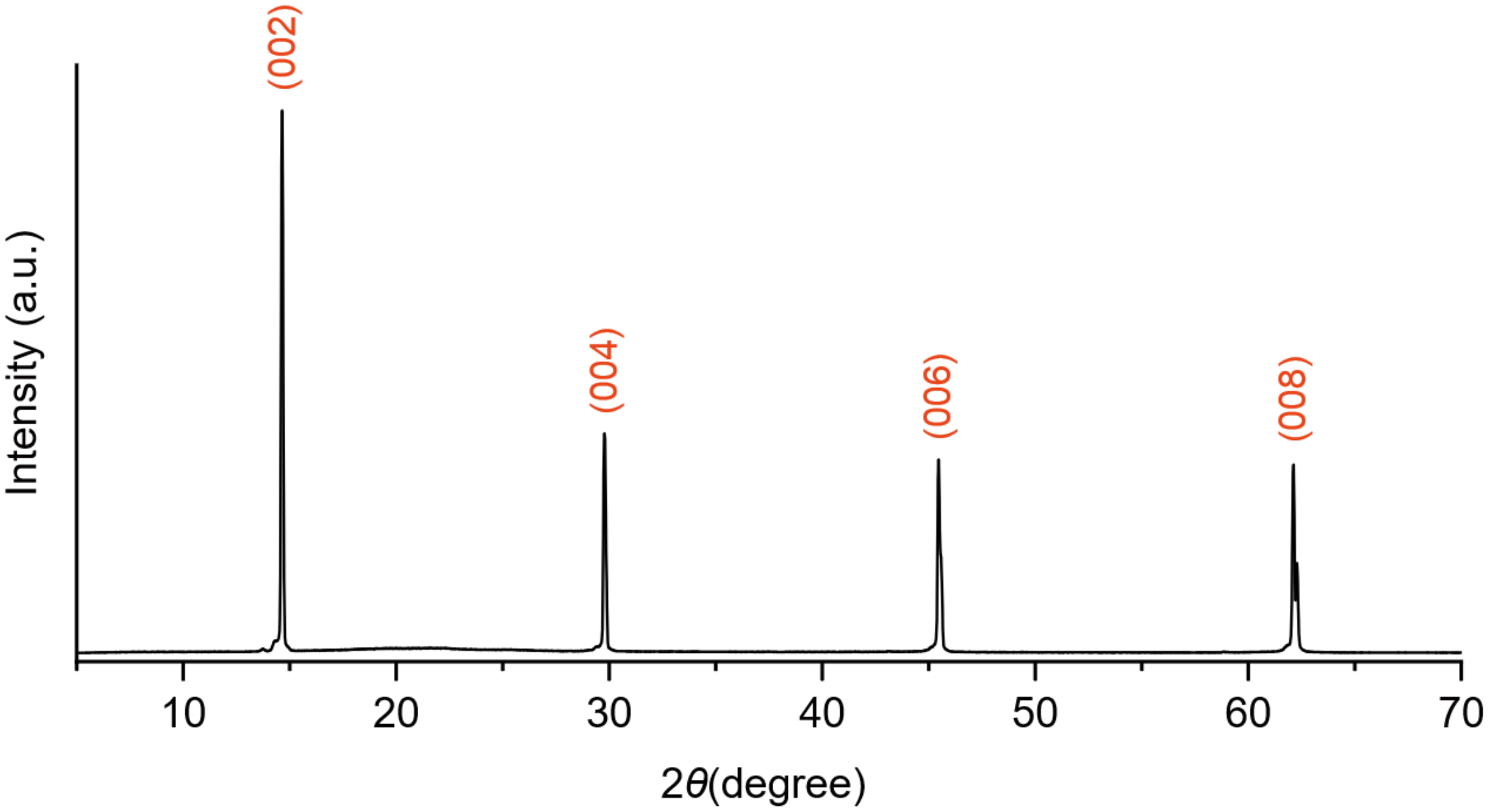

**Figure S1.** X-ray diffraction (XRD) pattern of our $Co_{1/3}TaS_2$ single crystal. XRD measurement was performed at room temperature using Rigaku Miniflex II, with the crystal aligned along the (001) plane. The sharp peaks of (002n) with n=1, 2, 3, 4 confirm the high crystalline quality of our crystal.

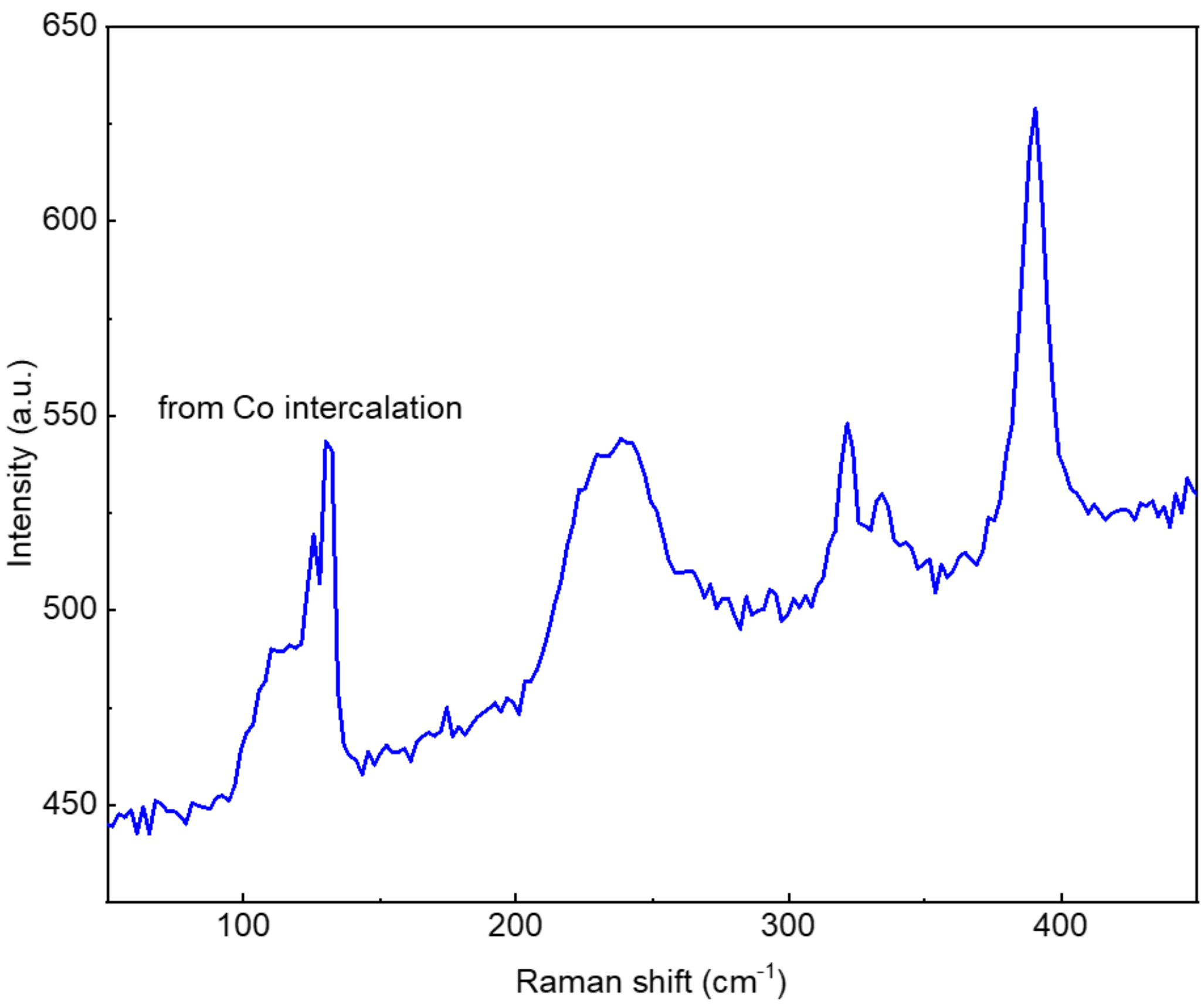

**Figure S2.** Raman spectra of a $Co_{1/3}TaS_2$ single crystal at room temperature. The Raman peak at around 137 $cm^{-1}$ originates from Co intercalation, as indicated, demonstrating well-arranged Co intercalation and the high quality of the crystals.

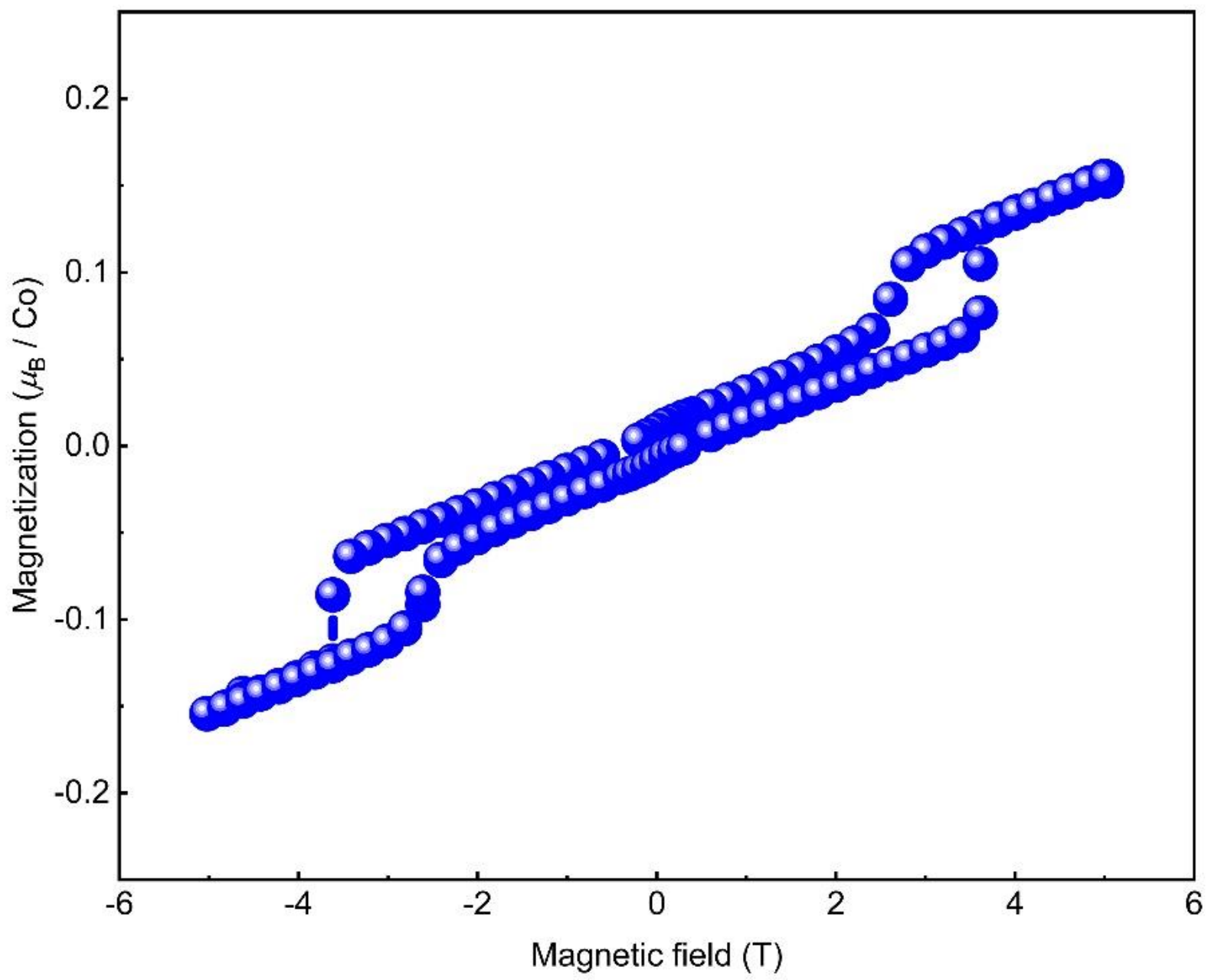

**Figure S3.** Magnetisation versus out-of-plane magnetic field ($M$-$H_z$ curve). The magnetic moment is small (0.1 μB/Co) at a high magnetic field and, most importantly, negligible at zero magnetic field, confirming the antiferromagnetic nature of the sample. These results are consistent with previously reported $M$-$H_z$ curves of $Co_{1/3}TaS_2$.

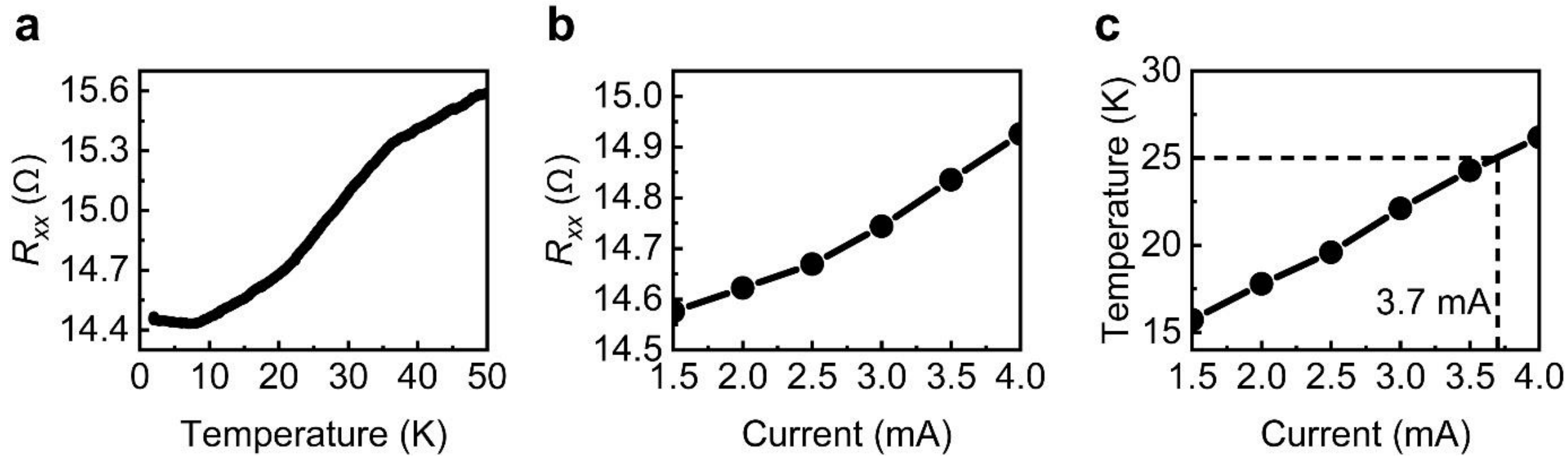

**Figure S4.** Quantification of the Joule heating effect. a) $R_{xx}$-$T$ curve of the $CoTaS_2$ device measured with a reading current of 0.1 mA. b) $R_{xx}$-$I$ curve of the $CoTaS_2$ device measured at 15 K. c) Estimated sample temperature from (a) and (b). A writing current exceeding 3.7 mA raises the sample temperature above $T_N$ of 25 K. Therefore, all switching currents were kept below 3.7 mA to ensure the device remained in the antiferromagnetic phase during the switching experiments.

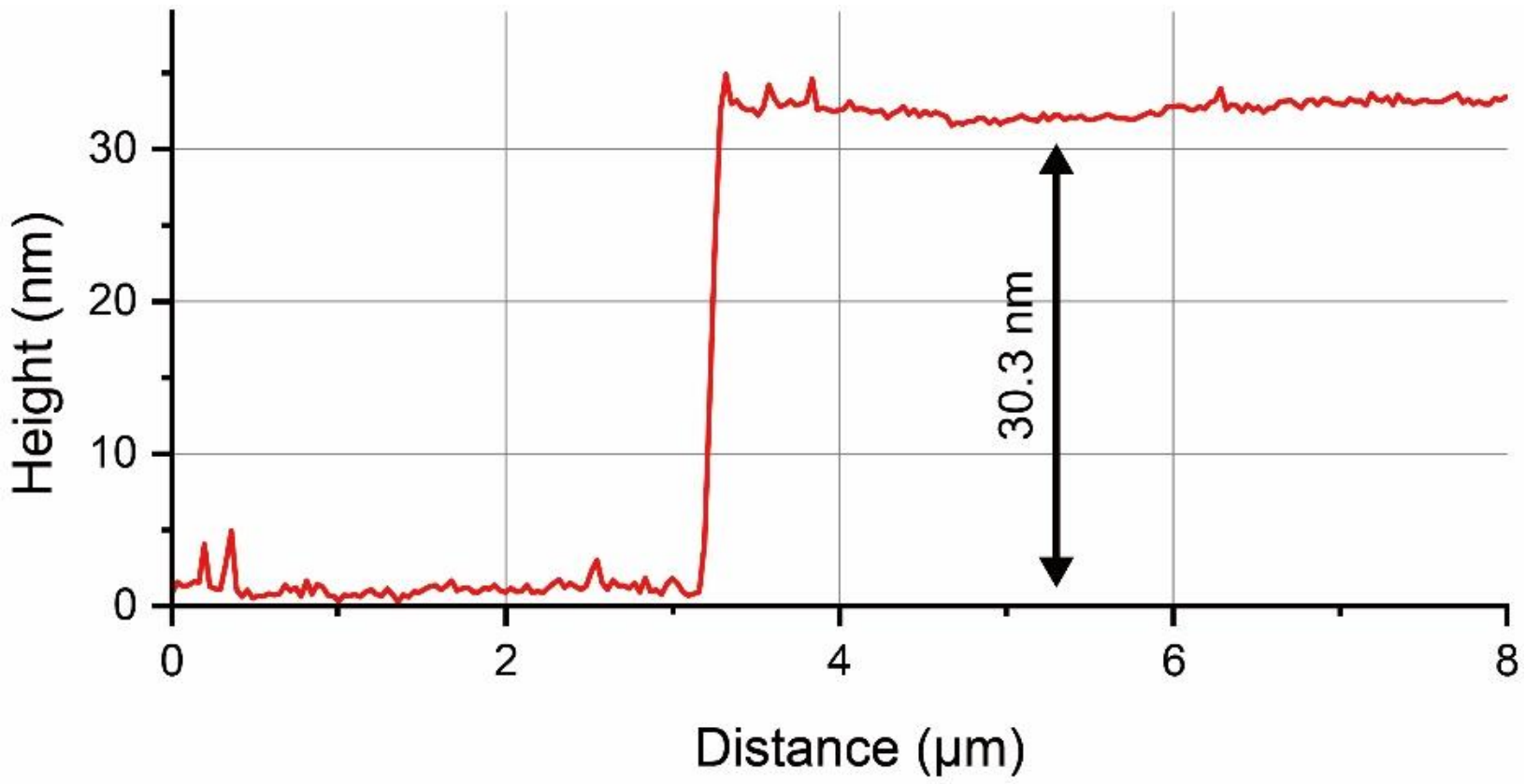

**Figure S5.** Thickness information of the $Co_{1/3}TaS_2$ device used for the experiments in Fig. 3. The thickness is measured to be about 30.3 nm by the atomic force microscopy.

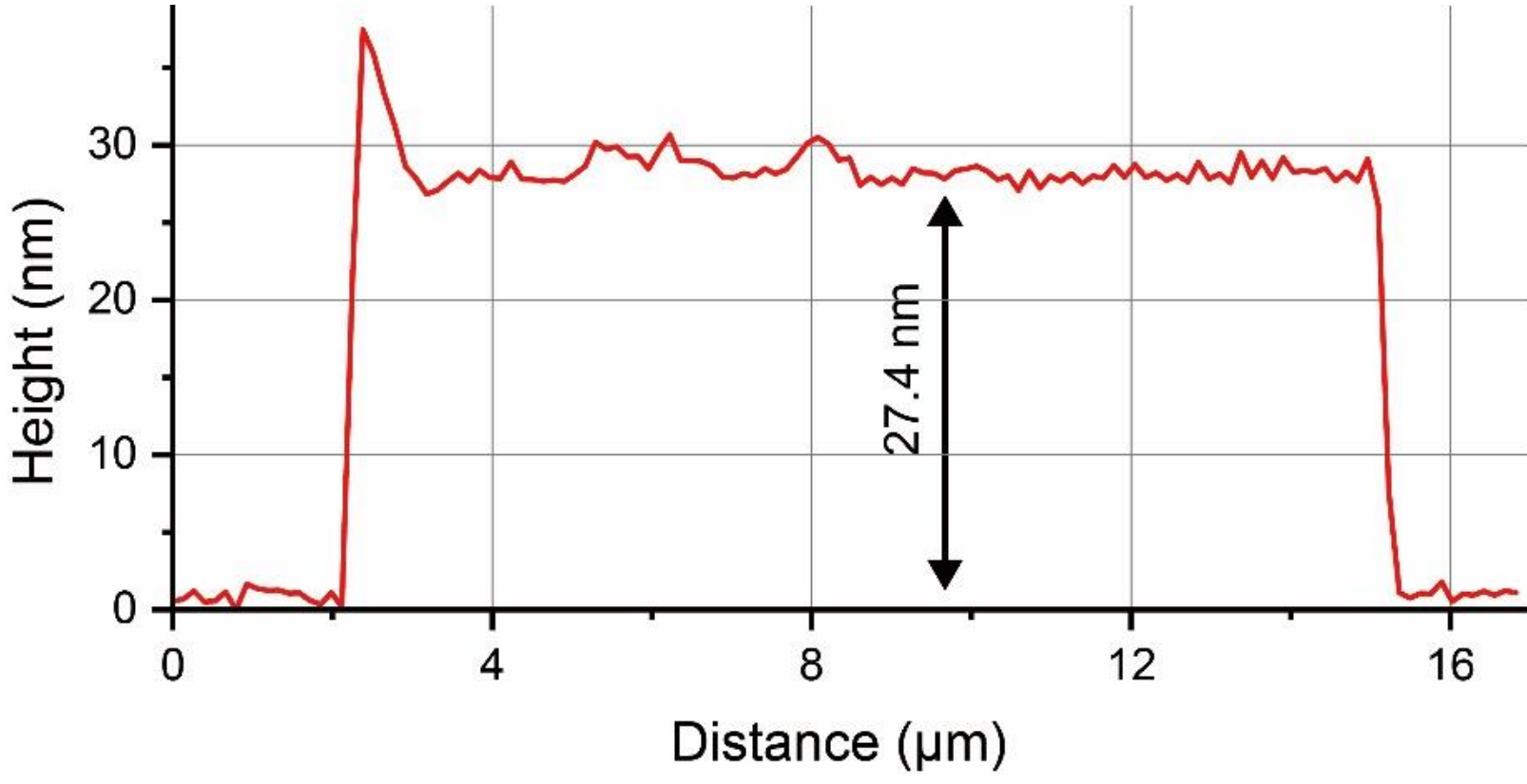

**Figure S6.** Thickness information of the $Co_{1/3}TaS_2$ device used for the experiments in Fig. 4. The thickness is measured to be about 27.4 nm by the atomic force microscopy.